\documentclass{aa}

\usepackage{graphicx}
\usepackage[colorlinks]{hyperref}

\usepackage{subcaption}
\hypersetup{citecolor={blue}}

\usepackage{txfonts}
\usepackage{multirow} 
\usepackage{booktabs}
\usepackage{placeins}

\begin{document}

   \title{The JWST-NIRCam view of Sagittarius C}
   \subtitle{III. The extinction curve}
   \titlerunning{The JWST-NIRCam view of Sagittarius C. III. The extinction curve}

   \author{Lucía Bravo Ferres
          \inst{1},
          Francisco Nogueras-Lara\inst{1,2},
          Rainer Schödel\inst{1}, Rubén Fedriani\inst{1}, Adam Ginsburg\inst{3}, Samuel Crowe\inst{4}, Jonathan C. Tan\inst{4,5}, Morten Andersen\inst{2}, Joseph Armstrong\inst{5}, Yu Cheng\inst{6}, Zhi-Yun Li\inst{4}
 }

   \authorrunning{Bravo Ferres, L., et al.}
   \institute{Instituto de Astrofísica de Andalucía, CSIC, Glorieta de la Astronomía            s/n, E-18008 Granada, Spain\\
              \email{lbravo@iaa.es}
         \and
             European Southern Observatory, Garching, Germany
        \and 
            Department of Astronomy, University of Florida, P.O. Box 112055, Gainesville, FL 32611
        \and
            Department of Astronomy, University of Virginia, Charlottesville, VA 22904, USA
        \and
            Department of Space, Earth \& Environment, Chalmers University of Technology, 412 93 Gothenburg, Sweden
        \and
            National Astronomical Observatory of Japan, 2-21-1 Osawa, Mitaka, Tokyo, 181-8588, Japan
}

   \date{Received; accepted}

  \abstract
   {Determining the infrared extinction curve towards the Galactic centre is crucial for accurately correcting observed data and deriving the underlying stellar populations. However, extinction curves reported in the literature often show discrepancies.}
   {We aim to derive the infrared extinction curve towards the Galactic centre based on JWST-NIRCam data for the first time, using observations of the Sagittarius C region in the 1–5\,$\mu$m range.}
   {We determined extinction ratios using two different methods, both based on measuring the reddening vector using the slope of red clump stars (whose intrinsic properties are well known) in observed colour-magnitude diagrams.}
   {The extinction curve derived in this work is in good agreement with previous results in the literature. We obtained the following extinction ratios relative to F162M: $A_\mathrm{F115W} : A_\mathrm{F162M} : A_\mathrm{F182M} : A_\mathrm{F212N} : A_\mathrm{F360M} : A_\mathrm{F405N} : A_\mathrm{F470N} : A_\mathrm{F480M} = 1.84 \pm 0.03 : 1.00 : 0.789 \pm 0.005 : 0.607 \pm 0.014 : 0.306 \pm 0.011 : 0.248 \pm 0.017 : 0.240 \pm 0.019 : 0.21 \pm 0.03$. Additionally, we found different values of the extinction index for the short- ($\lambda \sim 1-2.5\,\mu$m, $\alpha \sim 2$) and long-wavelength ($\lambda \sim 2.5-5\,\mu$m, $\alpha \sim 1.4$) regimes, with the extinction curve flattening at longer wavelengths. Comparison with extinction curves derived both inside and outside the Galactic centre suggests that the infrared extinction curve does not significantly vary in the central regions, and shows no significant evidence of variations between different lines of sight beyond the inner Galaxy within the uncertainties.}
    {}


   \maketitle

\section{Introduction} \label{intro}

    The Galactic centre (GC) represents a unique astrophysical laboratory, as it is the closest galactic nucleus and the only one where individual stars can be resolved down to milliparsec scales \citep[e.g.][]{Genzel2010}. However, studying its stellar content and structure is challenging due to the extreme stellar crowding and high interstellar extinction, which can exceed $A_V \gtrsim 30$~mag \citep[e.g.][]{Scoville2003, Nishiyama2008, Fritz2011, Schoedel2010}. This severe extinction restricts observations to the infrared regime, where high-angular-resolution photometry is essential to mitigate crowding effects \citep[e.g.][]{NoguerasLara2018}. Although intrinsic stellar colour values in the near-infrared are relatively small \citep[e.g.][]{Schoedel2010, NoguerasLara2018}, their variation across spectral types complicates photometric classification \citep[e.g.][]{Schoedel2014}. 

    An additional difficulty is that different wavelength dependences of the extinction have been reported in the literature \citep[e.g][]{Nishiyama2006,Schoedel2010,Fritz2011,NoguerasLara2019a}. It was generally assumed that the extinction curve in the $\sim 1-2.5\,\mu$m range follows a power law of the form $A_{\lambda} \propto \lambda^{-\alpha}$ \citep[e.g.][]{Nishiyama2008, Fritz2011}, where $\lambda$ is the wavelength and $\alpha$ is the extinction index. Previous studies found that $\alpha > 2$ \citep[e.g.][]{Nishiyama2006, Schoedel2010, AlonsoGarcia2017, NoguerasLara2018}, although more recent work shows that the extinction index varies with wavelength \citep[e.g.][]{Hosek2018, NoguerasLara2019a, NoguerasLara2020}. One of the most detailed spectroscopic studies of the extinction curve towards the GC was presented by \citet{Fritz2011}, who used hydrogen emission lines from the Short Wave Spectrometer of the Infrared Space Observatory \citep[ISO-SWS;][]{Graauw1996} and spectroscopy from the Spectrograph for INtegral Field Observations in the Near Infrared \citep[SINFONI;][]{Eisenhauer2003}, an integral-field spectrograph mounted on the Very Large Telescope (VLT). While they found an approximate power law in the $\sim 1.2-2.5\,\mu$m, the extinction curve longwards of $\lambda\sim2.5\,\mu$m appears to be complex \citep[e.g.][]{Lutz1999}.
    
    An accurate determination of the extinction curve requires studying the spectral energy distribution of stars with known properties. Ideally, this should be done using narrow-band filters and across a broad wavelength range to avoid non-linear photometric effects associated with broad-band filters \citep[for further details see][]{Jones1980, Straizys2008, MaizApellaniz2020, NoguerasLara2021a}. This makes the Near-Infrared Camera (NIRCam; \citeauthor{Rieke2005} \citeyear{Rieke2005}) on board the \textit{James Webb} Space Telescope (JWST) an ideal instrument for extinction studies as it combines high angular resolution with multi-filter coverage from 1 to 5\,$\mu$m.

    In this context, the GC Sagittarius C (Sgr\,C) region was recently observed with NIRCam \citep{Crowe2025, Bally2025}. Sgr\,C is a star-forming region that hosts ionised (HII) regions; it is located on the western edge of the Milky Way’s nuclear stellar disc \citep[e.g.][]{Lang2010, Kendrew2013}. Embedded in a dense molecular cloud, it lies in a region of high extinction \citep[$A_\mathrm{K_s} = 2.56\pm0.36$\,mag;][]{NoguerasLara2024}, making it an excellent target for deriving the extinction curve \citep[e.g.][]{Henshaw2023}. In this study we used the NIRCam observations of Sgr\,C to derive the extinction curve in the wavelength range 1.15 to 4.8\,$\mu$m.

\section{Observations and data reduction} \label{data}

    Observations of Sgr\,C were conducted with JWST-NIRCam on 22 September 2023 (programme ID: 4147, PI: S. Crowe). NIRCam consists of two redundant modules (A and B) separated by a central gap of $44''$, each providing a field of view of $2.2' \times 2.2'$ and containing both a short-wavelength (SW; $0.6$–$2.3\,\mu$m) and a long-wavelength (LW; $2.4$–$5.0\,\mu$m) channel \citep{Burriesci2005}. Each module contains four SW detectors with inter-detector gaps of $4''$–$5''$ and a pixel scale of $0.031''$/pixel, along with two LW detectors with a pixel scale of $0.061''$/pixel \citep{Burriesci2005}. To ensure full spatial coverage and mitigate detector gaps, six dither positions were employed using the FULLBOX 6TIGHT dither pattern, resulting in a final effective field of view of approximately $2' \times 6'$. The observations used four SW filters (F115W, F162M, F182M, and F212N) and four LW filters (F360M, F405N, F470N, and F480M).

\begin{figure*}[t]
        \centering
        \includegraphics[width=0.9\linewidth]{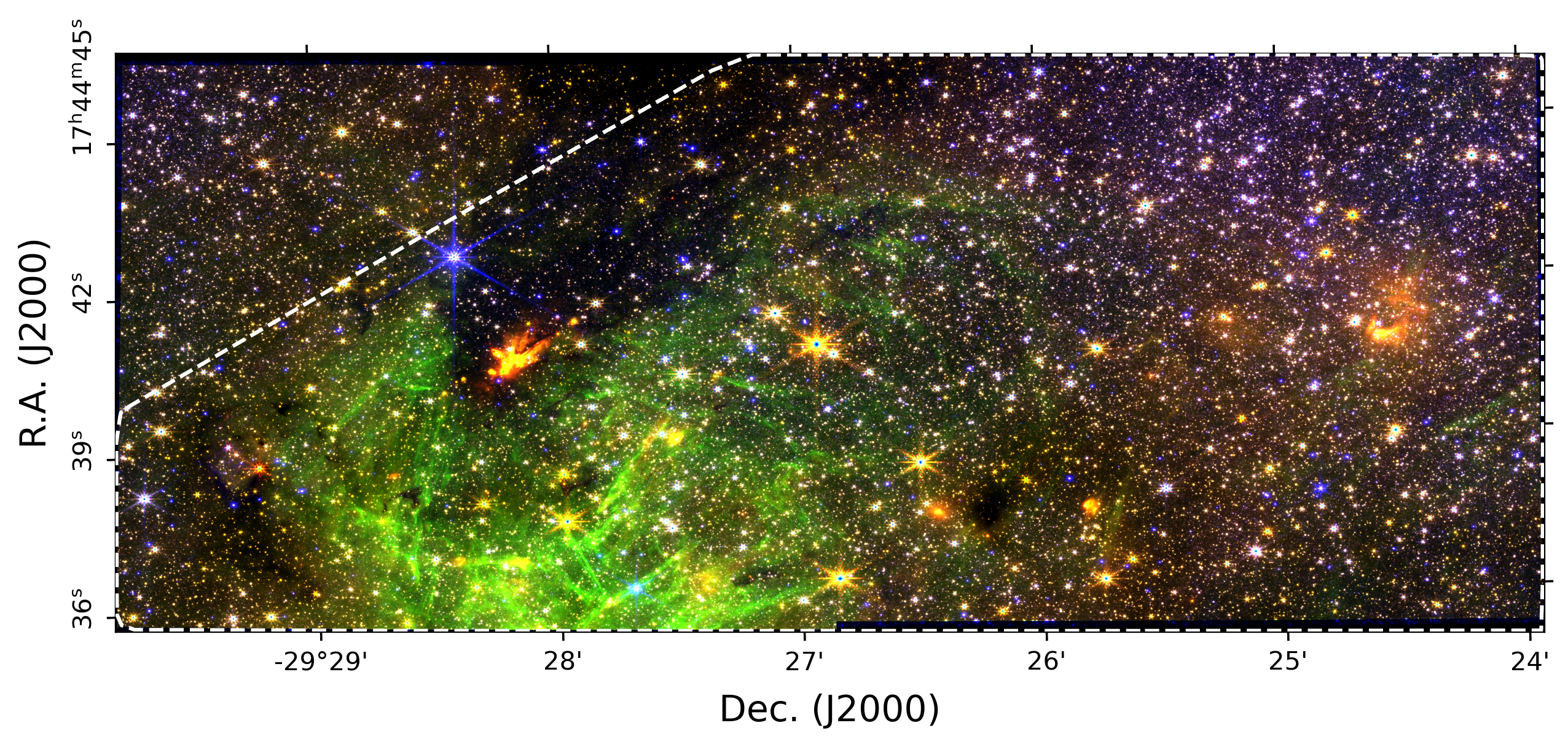}
        \caption{False-colour red-green-blue image of the Sgr\,C region created from Stage\,3 mosaics. The red, green, and blue channels correspond to the F480M, F405N, and F162M filters, respectively. The dashed white line outlines the region overlapping with the GALACTICNUCLEUS survey (see Sect.\,\ref{GNS}).}
        \label{mosaic_ex}
    \end{figure*}

    The data were reduced using the JWST standard pipeline (\texttt{jwst} v\texttt{1.12.5}; \citeauthor{Bushouse2023} \citeyear{Bushouse2023}), which includes detector corrections (Stage\,1), calibrations (Stage\,2), and image combination (Stage\,3). Figure\,\ref{mosaic_ex} shows a false-colour red-green-blue image of Sgr\,C created from Stage 3 mosaics. Residual $1/f$ noise, not fully addressed by the pipeline, was corrected using \texttt{image1overf.py}\footnote{\url{https://github.com/chriswillott/jwst/blob/master/image1overf.py}}, following the method described in \citet{Crowe2025}. Throughout this work we used Stage\,2 images with $1/f$ noise removed, because the photometric zero point of each NIRCam detector is different. We used the corresponding Vega magnitude zero points provided by the STScI\footnote{\texttt{NRC\textunderscore ZPs\textunderscore 1126pmap.txt}, to be found in \url{https://jwst-docs.stsci.edu/jwst-near-infrared-camera/nircam-performance/nircam-absolute-flux-calibration-and-zeropoints}.}. For further details on the filter set, observational setup and data reduction of the Sgr\,C NIRCam data, we refer to \cite{Crowe2025}.

\section{Photometric catalogues} 

\subsection{Point spread function fitting photometry}

We carried out point spread function (PSF) fitting photometry on the Stage\,2 NIRCam images, i.e. before combining them into mosaics. This approach was chosen because the photometric zero points vary significantly between different NIRCam detectors. This also allowed us to estimate robust uncertainties from independent measurements because the majority of sources were observed multiple times, that is, in images with different dither offsets and even with different detectors. 

We extracted the image, its corresponding error map, and the pixel map with the data quality flags from the multi-extension fits files. The error map is automatically generated by the JWST Stage\,2 pipeline and contains the uncertainty estimates for each pixel, provided as standard deviations. These uncertainties are computed by combining the variance due to Poisson noise and read noise, which are also included as separate extensions in the FITS file. We created saturation masks for each image by using the corresponding pixel data quality flags. A bad pixel map was created from the non-saturated pixels that were flagged as hot, cold, or `DO\_NOT\_USE'. We filled all bad and saturated pixels in the error map with the median of the ten highest valid values of the error map, which was necessary for subsequent processing. The precise value used to fill the error map did not have any significant effect on the outcome.

Subsequently, we used the {\it stpsf}\footnote{\url{https://stpsf.readthedocs.io/en/latest/}} tool to create theoretical PSFs for each detector and filter. Saturated stars were repaired by fitting the theoretical PSFs to their unsaturated wings, after having subtracted bright, but unsaturated stars from their vicinity. For this, we used a modified version of the \texttt{repair\_saturated} routine from the {\it StarFinder} software package \citep{Diolaiti2000}. Repairing the saturated stars proved to be essential to minimise the photometric contamination of nearby stars and to avoid the detection of spurious sources in the diffraction spikes of the saturated stars.

Finally, we used {\it StarFinder} to carry out PSF fitting photometry and astrometry, because it is optimised for crowded fields. We modified {\it StarFinder} to always assume positive extended emission throughout the image, because this makes physical sense for NIRCam in this extremely bright field. The correlation threshold was set to 0.85, the detection threshold to 3$\,\sigma$ with two iterations, the background fitting box to 11 pixels and the {\it deblend} parameter was set to 1. A correlation threshold of 0.85 is stricter than the standard 0.7 of \textit{StarFinder}, but well justified given the very well-defined and stable PSF of NIRCam \citep[see e.g.][]{Schoedel2010a, GallegoCano2018}. After tests varying these parameters, we found that they provided the optimum trade-off between high completeness, avoidance of detecting spurious sources, and speed in the source detection process. As concerns the background box, it can be varied without any significant impact on the results. The previously mentioned error map was used as an uncertainty map for the procedure. Finally, we used the NIRCam zero points of each filter to convert from calibrated fluxes to magnitudes.

\subsection{Final catalogues} \label{subsec:final_lists}

Once we obtained the star lists for all Stage\,2 images, we merged them to create the final catalogues for each photometric filter. Before this step, we aligned each list using a set of reference stars selected from the final F182M list, to ensure that all catalogues were in the same coordinate system. This is crucial for the cross-matching process in the next step. We chose the F182M filter as a reference because the saturation effect is less severe in the SW channel, and the positional offsets between common stars are lower in this filter compared to the other SW filters. Reference stars were selected with magnitudes in the range $14 < \mathrm{F182M} < 18$\,mag, avoiding both saturated sources and faint stars with large photometric uncertainties. We also verified that the selected reference stars were homogeneously distributed across the field, rather than concentrated in a specific region, to ensure a robust alignment over the full image area.

We built the final catalogues using the function \texttt{match\_to\_catalog\_sky} from the Python package \texttt{Astropy} \citep[version \texttt{5.2.2};][]{AstropyCollaboration2013, AstropyCollaboration2018, AstropyCollaboration2022} to identify common stars in the overlap regions between the detectors. For each common star, we computed the mean value of parameters such as its magnitude and coordinates. We chose a match radius of $0.05''$, which is smaller than the diffraction limit for all filters, but large enough to accommodate also fainter stars that show greater astrometric scatter (see Sect.\,\ref{data}). Given that systematic astrometric offsets between chips are below $0.01''$, the expected astrometric uncertainties are significantly smaller than our matching threshold. Therefore, a search radius of $0.05''$ ensures robust cross-matching without introducing spurious matches.

To align the chip-by-chip lists for each filter (i.e. those obtained from the Stage\,2 images) with the F182M reference catalogue, we first cross-matched them using a radius of $0.05''$. We then applied sigma clipping, keeping only the stars whose positional offsets fell within 3$\,\sigma$ of the distribution of offsets among the matched sources. We then computed the mean offsets in right ascension ($\alpha$) and declination ($\delta$) and subtracted them from the stellar coordinates. Finally, we merged the aligned lists using the same search radius ($0.05''$) to obtain the final catalogues for the remaining filters.                
    
\subsection{Photometric uncertainties}

In the final catalogues, some stars are detected only once, while others have multiple detections. In the latter case, for each detection, we first derived the flux from the computed magnitude. Then, we estimated the flux uncertainty, $\Delta f$ following a similar approach to that of \cite{NoguerasLara2018}:

\begin{equation}
    \Delta f = \frac{f_{max}-f_{min}}{2\sqrt{N}}
    \label{phot_unc}
,\end{equation}

\noindent where $f_{max}$ and $f_{min}$ are the largest and the lowest values of the flux for the star, respectively, and $N$ is the number of detections of the star. 

Next, we converted the mean flux $\overline{f}$ and its uncertainty $\Delta f$ into magnitudes by computing the magnitudes corresponding to $\overline{f} + \Delta f$ and $\overline{f} - \Delta f$. We then estimated the photometric uncertainty in magnitudes as the average of the absolute differences between the magnitude corresponding to $\overline{f}$ and those corresponding to $\overline{f} \pm \Delta f$.

\begin{figure}
\centering
    \includegraphics[width=0.8\linewidth]{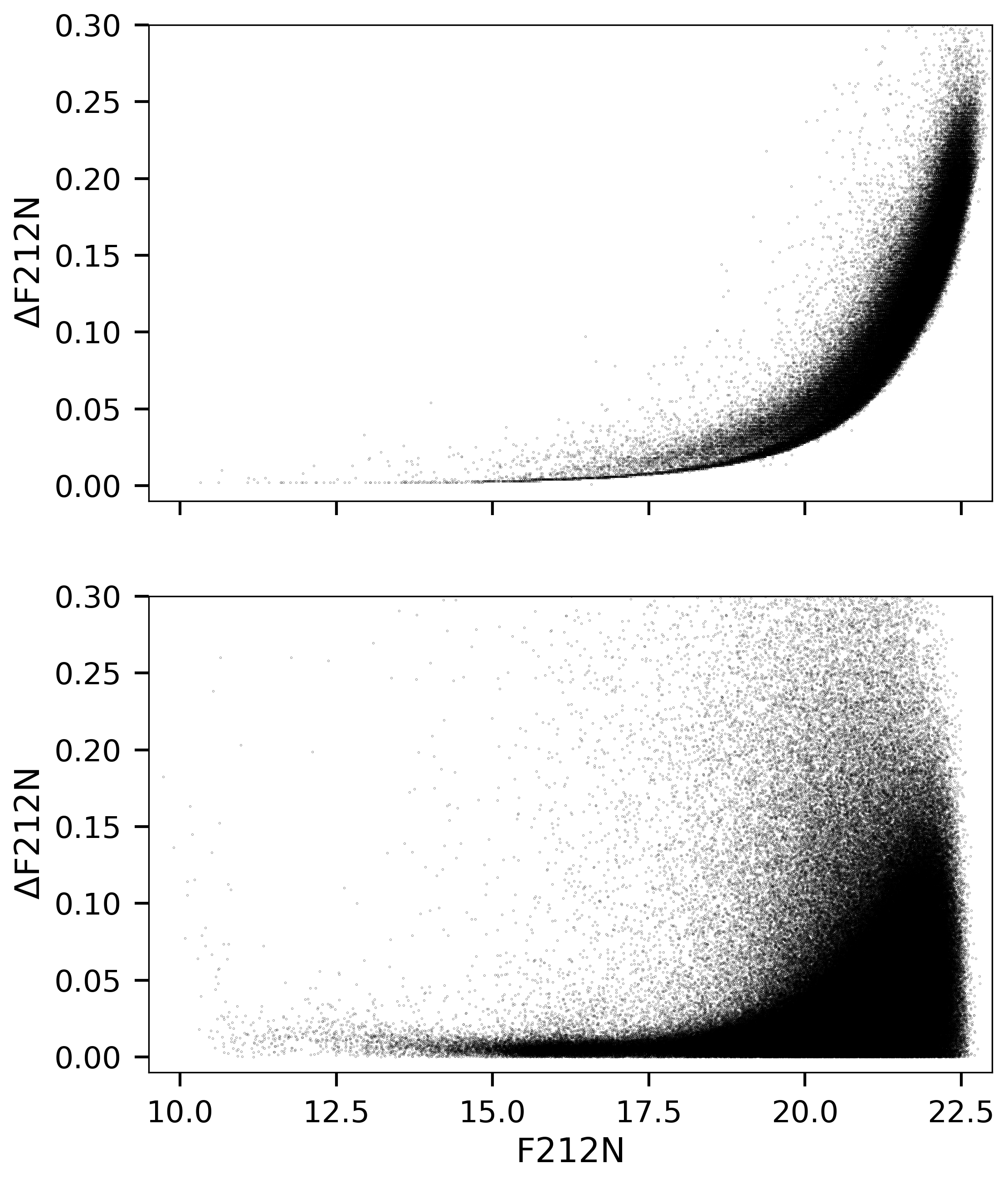}
    \caption{Top: Uncertainties returned by \emph{StarFinder}. Bottom: Uncertainties estimated from multiple detections of the same stars.}
    \label{Phot_unc}
\end{figure}

We used the uncertainties provided by the \emph{StarFinder} package, which fits the stars with an IDL implementation of the Newton-Gauss method and uses the provided pixel uncertainty map to formally estimate the uncertainties. Stars with lower signal-to-noise ratios thus have larger uncertainties (Fig.\,\ref{Phot_unc} top). Since this procedure depends on providing an accurate pixel uncertainty map, we cross-checked the uncertainties by using the measurements of stars with multiple detections. As the bottom panel of Fig.\,\ref{Phot_unc} shows, the photometric uncertainties from multiply detected stars can be very small due to the random nature of this method (e.g.\ when two measurements are very similar), but the upper envelope of the cloud of uncertainties is in good agreement with the uncertainties provide by \emph{StarFinder}, thus demonstrating the validity of the uncertainties delivered by \emph{StarFinder}.

\subsection{Colour-magnitude diagrams} \label{cmds_sect}

\begin{figure}
    \centering
    \includegraphics[width=1\linewidth]{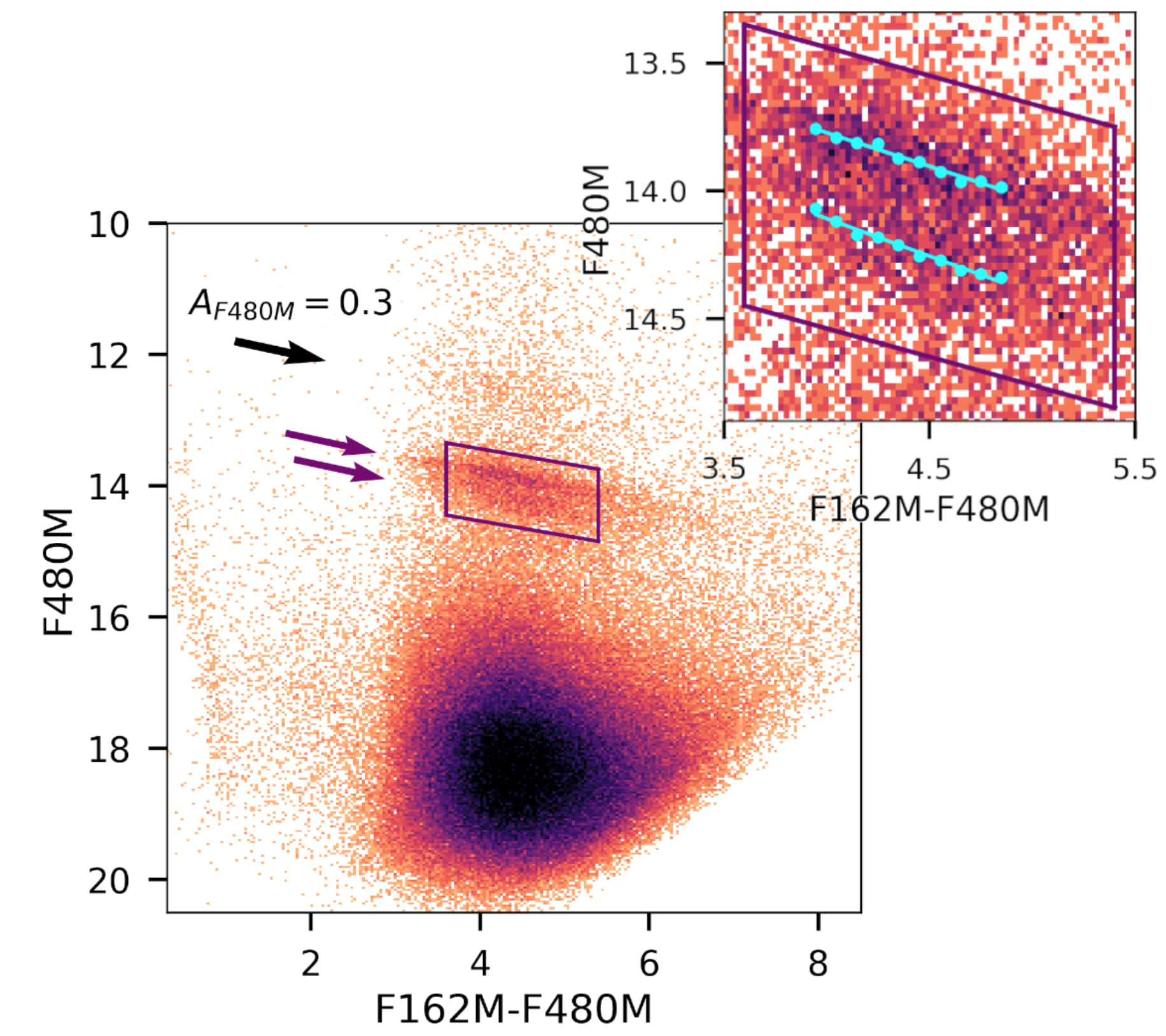}
    \caption{F480M versus F162M-F480M CMD. The purple box highlights the RC feature, and the black arrow is the reddening vector for $A_\mathrm{F480M} = 0.3$\,mag. The two purple arrows indicate the bright (top) and faint (bottom) RC features. The inset shows a zoomed-in view of the RC region, where the two features are identified using the GMM approach described in Sect.\,\ref{subsec:methods}. The cyan points correspond to the positions of the RC features in each bin, and the cyan lines show the best-fitting linear regressions. The uncertainties, although included, are smaller than the marker size and therefore not visible. The purple box in the CMD is also included for reference.}
    \label{CMD}
\end{figure}

To build the colour-magnitude diagrams (CMDs), we cross-matched the catalogues from selected pairs of filters using a distance of $0.05''$. Figure\,\ref{CMD} shows the F480M versus F162M-F480M CMD. The purple box indicates the red clump (RC) feature, identified as an overdensity in the diagram. It consists of giant stars burning helium in their cores \citep{Girardi2016}. Due to their weak dependence on age and metallicity, the intrinsic properties of these stars, which are well characterised, make them ideal for tracing extinction and determining distances. Moreover, since the RC is a prominent feature in the CMD, we can see clearly how it extends parallel to the reddening vector (black arrow in Fig.\,\ref{CMD}). This vector was computed using the extinction curve derived in this work (see Sect.\,\ref{subsec:methods}). 

We also observe the presence of a bright and a faint RC. This double RC feature has been interpreted as evidence of two stellar populations formed in separate star formation episodes—an older component ($\sim$8\,Gyr) and a younger one ($\sim$1\,Gyr; see \citealt{NoguerasLara2019a}). Additionally, we considered stars with F162M-F480M $\lessapprox 3$\,mag to belong to the foreground population \citep[e.g.][]{NoguerasLara2021}.

\section{Extinction curve}

We computed the extinction curve using two methods, both based on deriving extinction ratios from the slope of the RC features, as described in Sect.\,\ref{subsec:methods}. The first, referred to as the reference filter method, uses F162M as the reference; extinction ratios are computed relative to this filter. The second, the consecutive filters method, derives extinction ratios between adjacent filters in wavelength. 

Ideally, we would use the shortest-wavelength filter, F115W, as the baseline in the reference filter method. Colour combinations involving F115W span a broader wavelength range and extend further into the blue part of the spectrum, which is more sensitive to extinction. This makes the RC feature appear more extended in CMDs including F115W. However, because of the high extinction in this filter, the number of stars detected is very small ($\sim3$ times fewer than in F470N and $\sim18$ times fewer than in F182M), leading to poorly populated CMDs, which can affect the reliability of the method. Therefore, we chose F162M, as it is the second bluest option in our filter set. 

\subsection{Reference filter method} \label{subsec:methods}

The slope of the RC features can be used to derive extinction ratios, as it corresponds to $A_{\lambda_1}/E_{\lambda_2-\lambda_1}$ \citep{Nishiyama2006}, where $A_{\lambda_1}$ is the extinction in magnitudes at a given wavelength $\lambda_1$, and $E_{\lambda_2-\lambda_1}$ is the colour excess, defined as $E_{\lambda_2-\lambda_1} = A_{\lambda_2}-A_{\lambda_1}$. From this relation, the extinction ratio can be expressed as 

\begin{equation}
    \frac{A_{\lambda_2}}{A_{\lambda_1}} = 1 + \frac{1}{s}
    \label{extrat}
,\end{equation}

\noindent where $s$ is the slope of the RC in the $\lambda_1$ versus $\lambda_2 - \lambda_1$ CMD. Thus, for the reference filter method, we adopted $\lambda_2 = \mathrm{F162M}$. To derive the slope of the RC features, we applied the methodology described in \cite{NoguerasLara2019} and \cite{NoguerasLara2020}. Namely, we divided the RC region into five to seven narrow vertical bins, depending on the colour width of the RC features. We show the selection boxes and binning used for each CMD in the reference filter method in Fig.\,\ref{app_method1}. The selection boxes were defined to ensure a relatively homogeneous density of RC stars across the CMDs, minimising the impact of incompleteness at the faint end. We then verified the presence of a double RC sequence (see Sect.\,\ref{cmds_sect}) by analysing the magnitude distribution within each bin, fitting both a single-Gaussian and a two-Gaussian model using the \texttt{GaussianMixture} (GMM) function from the SCIKIT-LEARN Python package \citep{Pedregosa2011}. In all cases, we found that the two-Gaussian model describes best the underlying distribution, as determined by the Bayesian information criterion \citep{Schwarz1978} and the Akaike information criterion \citep{Akaike1974}. 
     
For each bin, we calculated the representative colour as the median of the stellar colours within the bin, and estimated its uncertainty as the standard error of the median ($1.25\sigma/\sqrt{N}$). The peaks of the Gaussians identified by the GMM provide the y-axis positions of the RC features—that is, the magnitude values corresponding to the bright and faint RC populations in each bin. The uncertainty was computed as the standard error of the mean of each Gaussian distribution. The resulting points and their uncertainties are shown in Fig.\,\ref{CMD} for the F480M versus F162M-F480M CMD as an example. The Gaussian fit results for the remaining CMDs are shown in Appendix\,\ref{app1}.

To derive the slopes of the RC features and their associated statistical uncertainties, we applied a jackknife resampling algorithm. Specifically, we determined the slope $N$ times ($N$ being the number of bins), dropping the value of a different bin each time. We computed the final slope as the mean of the resulting values, and the statistical uncertainty as their standard deviation. We also tested different sources of systematic uncertainties:

\begin{itemize}
    \item The cross-match radius (defined as the maximum allowed positional difference between two stars detected in different filters for them to be considered the same source): $0.01''$ and $0.1''$. 
    \item The number of bins (and consequently their width): we tested using two fewer bins than in the nominal case (e.g. using four if it was six), as well as ten bins in all cases.
    \item The colour determination: we tried the mean of the distribution and the central colour of the bin.
    \item The exclusion of the last bin, which typically contains fewer stars and is more affected by incompleteness.
    \item The RC selection box along both the magnitude and colour axes: we used a smaller and a bigger box. 
\end{itemize}

The test on the cross-match radius also accounts for the different angular resolutions between the SW and LW channels; however, this test had a negligible effect on the results. Variations in the RC box along the colour axis further helped us evaluate completeness effects: larger boxes include the less dense—and therefore less complete—edges of the RC, while smaller boxes focus on its densest, most complete region. Table\,\ref{systematics} shows the individual contributions to the systematic uncertainty for each filter and for both the bright and faint RC features. The total systematic uncertainty was computed as the quadratic sum of all components. The variation of the RC selection box yields the largest contribution to the total systematic uncertainty, as shown in Table\,\ref{systematics}.

Finally, we derived the extinction ratios for each RC feature from Eq.\,\ref{extrat} and calculated the statistical and systematic uncertainties from the uncertainties of the slopes. Tables\,\ref{results_short} and \ref{results_long} present the resulting extinction ratios and their associated uncertainties for the SW and LW filters, respectively. The values are shown separately for the bright and faint RC features, along with the measured slopes. We also report the average extinction ratios, with their statistical and systematic uncertainties computed via quadratic propagation. 
    
\begin{figure}
    \centering
    \includegraphics[width=1\linewidth]{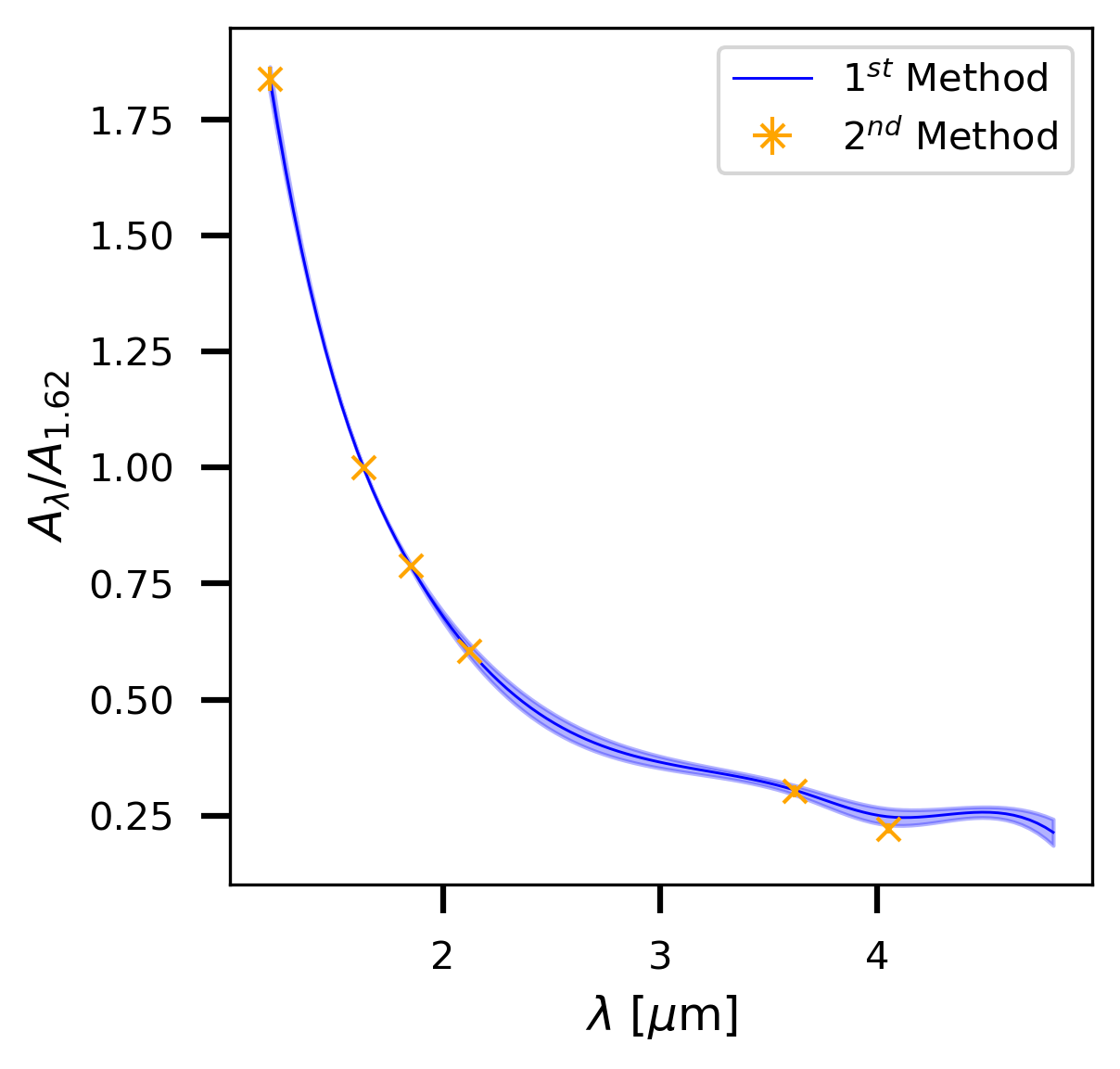}
    \caption{Extinction curve derived using the reference (blue, first method) and the consecutive filter (orange, second method) methods with their uncertainties. The x-axis shows the effective wavelength for each filter. The extinction curve from the first method has been spline-interpolated (see Sect.\,\ref{subsec:methods}). In both cases, the plotted extinction ratios correspond to the mean values obtained from the bright and faint RC features. In most cases, the uncertainties are smaller than the marker size and thus not visible.}
    \label{resultboth}
\end{figure}

Figure \ref{resultboth} shows the final results. We adopted the final extinction ratio values as the mean of those obtained from the bright and faint RC features. For each RC, we computed the total uncertainty by adding in quadrature the statistical and systematic components. The uncertainty of the mean extinction ratio was then derived by propagating the statistical and systematic uncertainties quadratically.  
    
In addition, we interpolated the extinction curve derived with the reference filter method, along with its associated uncertainties, using the \texttt{CubicSpline} function from the SciPy library \citep{Virtanen2020}. For each filter, we derived the effective wavelength, $\lambda_{\mathrm{eff}}$, as described in Appendix\,\ref{eff_wls_sect}. As part of this process, we also computed the extinction index between each filter and the reference filter. The resulting values are shown in Tables\,\ref{results_short} and \ref{results_long} for both the bright and faint RC, along with their average.

\renewcommand{\arraystretch}{1.3}
\setlength{\tabcolsep}{5pt}

\begin{table}[htbp]
\centering
\caption{Systematic uncertainties of the slope for each filter and RC feature.}
\label{systematics}
\small
\begin{tabular}{llccccc}
\hline\hline
Filter &  & $r$ & \# bins & Colour & Last bin & RC box \\
\hline
\multirow{2}{*}{F115W}  & Bright & 0.0003 & 0.017 & 0.005 & 0.008 & 0.015 \\
       & Faint  & 0.0007 & 0.005 & 0.005 & 0.023 & 0.03 \\
\addlinespace[1ex]
\multirow{2}{*}{F182M}  & Bright & 0.005  & 0.05 & 0.03 & 0.007 & 0.23 \\
       & Faint  & 0.017  & 0.10 & 0.04 & 0.014 & 0.23 \\
\addlinespace[1ex]
\multirow{2}{*}{F212N}  & Bright & 0.0023 & 0.011 & 0.016 & 0.007 & 0.11 \\
       & Faint  & 0.006 & 0.023 & 0.016 & 0.07 & 0.13 \\
\addlinespace[1ex]
\multirow{2}{*}{F360M}  & Bright & 0.004 & 0.006 & 0.005 & 0.008 & 0.022 \\
       & Faint  & 0.003 & 0.006 & 0.005 & 0.008 & 0.015 \\
\addlinespace[1ex]
\multirow{2}{*}{F405N}  & Bright & 0.0005 & 0.004 & 0.0016 & 0.007 & 0.021 \\
       & Faint  & 0.0016 & 0.007 & 0.0016 & 0.005 & 0.04 \\
\addlinespace[1ex]
\multirow{2}{*}{F470N}  & Bright & 0.0004 & 0.0018 & 0.013 & 0.007 & 0.013 \\
       & Faint  & 0.0004 & 0.004 & 0.03 & 0.010 & 0.05 \\
\addlinespace[1ex]
\multirow{2}{*}{F480M}  & Bright & 0.0004 & 0.0021 & 0.0014 & 0.003 & 0.03 \\
       & Faint  & 0.0015 & 0.0012 & 0.0015 & 0.0009 & 0.08 \\
\hline
\end{tabular}
\tablefoot{
The sources of uncertainty are: cross-match radius ($r$), number of bins (\# bins), method for determining the bin colour (colour), last bin removal (last bin), and the RC selection (box).
}
\end{table}

\renewcommand{\arraystretch}{1.3}
\setlength{\tabcolsep}{6pt}

\begin{table*}[htbp]

\centering
\caption{Slopes, extinction ratios ($A_\mathrm{\lambda}/A_\mathrm{F162M}$), and extinction indices ($\alpha_{\lambda-\mathrm{F162M}}$) for SW filters.}
\label{results_short}

\begin{tabular}{llccc}
\hline \hline
& & F115W & F182M & F212N \\
\hline
\multirow{2}{*}{Slope} 
  & Bright & $1.17 \pm 0.03 \pm 0.03$ & $3.72 \pm 0.04 \pm 0.15$ & $1.533 \pm 0.019 \pm 0.113$ \\
  & Faint  & $1.22 \pm 0.04 \pm 0.04$ & $3.74 \pm 0.08 \pm 0.10$ & $1.550 \pm 0.017 \pm 0.145$ \\
  \addlinespace[1.5ex]
\multirow{3}{*}{$A_\mathrm{\lambda} / A_{\mathrm{F162M}}$} 
  & Bright & $1.855 \pm 0.022 \pm 0.022$ & $0.7881 \pm 0.0018 \pm 0.0067$ & $0.605 \pm 0.003 \pm 0.017$ \\
  & Faint  & $1.82 \pm 0.03 \pm 0.03$ & $0.789 \pm 0.004 \pm 0.004$ & $0.608 \pm 0.003 \pm 0.023$ \\
  & Average & $1.838 \pm 0.019 \pm 0.019$ & $0.7885 \pm 0.0022 \pm 0.0040$ & $0.6065 \pm 0.0021 \pm 0.0143$ \\
  \addlinespace[1.5ex]
\multirow{3}{*}{$\alpha_{\mathrm{\lambda} - \mathrm{F162M}}$} 
  & Bright & $2.02 \pm 0.04 \pm 0.04$ & $1.915 \pm 0.018 \pm 0.071$ & $1.927 \pm 0.019 \pm 0.108$ \\
  & Faint  & $1.95 \pm 0.05 \pm 0.05$ & $1.90 \pm 0.04 \pm 0.04$ & $1.907 \pm 0.019 \pm 0.145$ \\
  & Average & $1.99 \pm 0.03 \pm 0.03$ & $1.908 \pm 0.022 \pm 0.041$ & $1.917 \pm 0.013 \pm 0.090$ \\
\hline
\end{tabular}
\tablefoot{ The values are given for the bright and faint RC features, as well as the average for the extinction ratios and indices. All quantities are shown along with their corresponding statistical and systematic uncertainties.}
\end{table*}

\begin{table*}[htbp]

\centering
\caption{Slopes, extinction ratios ($A_\mathrm{\lambda}/A_\mathrm{F162M}$), and extinction indices ($\alpha_{\mathrm{\lambda}-\mathrm{F162M}}$) for LW filters.}
\label{results_long}

\begin{tabular}{llcccc}
\hline \hline
& & F360M & F405N & F470N & F480M \\
\hline
\multirow{2}{*}{Slope}
  & Bright & $0.426 \pm 0.023 \pm 0.024$ & $0.325 \pm 0.017 \pm 0.023$ & $0.300 \pm 0.015 \pm 0.020$ & $0.263 \pm 0.008 \pm 0.028$ \\
  & Faint  & $0.460 \pm 0.036 \pm 0.019$ & $0.333 \pm 0.014 \pm 0.046$ & $0.33 \pm 0.03 \pm 0.06$     & $0.28 \pm 0.03 \pm 0.08$ \\
  \addlinespace[1.5ex]
\multirow{3}{*}{$A_\mathrm{\lambda} / A_{\mathrm{F162M}}$}
  & Bright & $0.299 \pm 0.011 \pm 0.012$ & $0.245 \pm 0.010 \pm 0.013$ & $0.231 \pm 0.009 \pm 0.012$ & $0.208 \pm 0.005 \pm 0.019$ \\
  & Faint  & $0.313 \pm 0.008 \pm 0.009$ & $0.250 \pm 0.008 \pm 0.028$ & $0.248 \pm 0.017 \pm 0.034$ & $0.220 \pm 0.018 \pm 0.049$ \\
  & Average & $0.306 \pm 0.007 \pm 0.008$ & $0.248 \pm 0.006 \pm 0.016$ & $0.240 \pm 0.010 \pm 0.016$ & $0.214 \pm 0.009 \pm 0.027$ \\
  \addlinespace[1.5ex]
\multirow{3}{*}{$\alpha_{\mathrm{\lambda} - \mathrm{F162M}}$}
  & Bright & $1.52 \pm 0.05 \pm 0.05$ & $1.55 \pm 0.05 \pm 0.06$ & $1.39 \pm 0.04 \pm 0.05$ & $1.455 \pm 0.022 \pm 0.090$ \\
  & Faint  & $1.46 \pm 0.03 \pm 0.04$ & $1.53 \pm 0.04 \pm 0.13$ & $1.32 \pm 0.07 \pm 0.11$ & $1.40 \pm 0.08 \pm 0.21$ \\
  & Average & $1.49 \pm 0.03 \pm 0.03$ & $1.54 \pm 0.03 \pm 0.07$ & $1.36 \pm 0.04 \pm 0.06$ & $1.43 \pm 0.04 \pm 0.11$ \\
\hline
\end{tabular}
\tablefoot{The values are given for the bright and faint RC features, as well as the average for the extinction ratios and indices. All quantities are shown along with their corresponding statistical and systematic uncertainties.}
\end{table*}

\subsection{Consecutive filters method}

The reference filter method always uses a fixed wavelength to estimate the extinction ratios. To check whether this approach misses any potential variation between consecutive wavelengths—and to validate the results obtained—we computed the extinction ratios between consecutive filters using the same approach described above. However, we did not derive the extinction ratios $A_\mathrm{F405N}/A_\mathrm{F470N}$ and $A_\mathrm{F470N}/A_\mathrm{F480M}$ because the differential extinction between these filter pairs is very small. This is not merely due to their proximity in wavelength, but rather because the extinction curve flattens significantly at longer wavelengths, leading to nearly constant extinction values (see Table\,\ref{results_long}). Consequently, the slope of the RC features in the F470N versus F405N–F470N and F480M versus F470N–F480M CMDs becomes excessively steep, with the RC feature exhibiting a very narrow spread in colour. In such cases, uncertainties in colour can dominate the observed slope, leading to unreliable estimates of slopes and thus extinction ratios. Moreover, the transmission curve of F480M overlaps with that of F470N. 

It is also important to note that in CMDs such as F480M versus F470N–F480M, which use only LW filters, the foreground population does not appear as a clearly separated component, unlike in other CMDs, but is instead blended with Sgr\,C stars. To address this, we identified and removed foreground stars using the F212N versus F162M–F212N CMD. This removal had a negligible impact on our results, as only a small fraction of these stars belong to the RC features (see Appendix\,\ref{app2}). 
    
The extinction curve derived using the consecutive filters method is shown in Fig.\,\ref{resultboth}. To enable a direct comparison, we converted the extinction ratios obtained with this approach to the reference filter system. As before, the final extinction-ratio values and their associated uncertainties were computed as the mean between the values obtained for the bright and faint RC features. As shown in Fig.\,\ref{resultboth}, the two methods yield extinction curves that agree within the uncertainties. 

Since the consecutive filters method does not allow us to derive extinction ratios for the two reddest filters, and given its agreement with the reference filter method, we adopt the extinction curve obtained with the reference filter method for the remainder of the discussion. 
   
\section{Discussion}

\subsection{Interpretation of the derived extinction curve}

Based on the results presented in Tables\,\ref{results_short} and \ref{results_long}, we find that the slopes—and consequently the extinction ratios—obtained from the faint RC are systematically larger than those from the bright RC. This leads to smaller values of the extinction index. However, the differences between the two components remain within the combined statistical and systematic uncertainties. We believe the small observed discrepancy is likely due to the secondary (faint) RC being intrinsically fainter and containing fewer stars than the bright RC. Therefore, it is more strongly affected by observational incompleteness and differential extinction. 

As shown in Fig.\,\ref{resultboth} and Tables\,\ref{results_short} and \ref{results_long}, our extinction curve flattens and appears to keep slightly descending in the LW regime. However, the ratio $A_\mathrm{F470N}/A_\mathrm{F162M}$ appears higher than the trend suggested by neighbouring filters in the LW regime. This deviation is visible as a bump in the interpolated extinction curve and is attributed to the strong CO ice absorption feature affecting the 4.7\,$\mu$m filter, as previously reported in Fig.\,7(a) of \citet[see also \citealt{Lutz1999, Moultaka2009}]{Ginsburg2023}.

We observe a clear difference in the extinction indices derived for the SW filters (see Table\,\ref{results_short}) and those for the LW filters (see Table\,\ref{results_long}). In the SW range, we obtain $\alpha_{\lambda-\mathrm{F162M}} \sim 2$, whereas in the LW range, the values decrease to $\alpha_{\lambda-\mathrm{F162M}} \sim 1.4$. This indicates a clear wavelength dependence of the extinction index when comparing the SW and LW regimes. However, within the SW and LW regimes our measurements are consistent with single $\alpha$-values. This contrasts with previous studies such as \citet{NoguerasLara2020} and \citet{Hosek2018}, which reported a noticeable wavelength dependence of the extinction index within the JH$\mathrm{K_s}$ bands. However, our uncertainties are likely too large to detect such a variation. 

\subsection{Comparison with the GALACTICNUCLEUS survey} \label{GNS}

To further assess our results using an independent dataset and to investigate a possible wavelength dependence of the extinction index in the $\sim 1.4$–$2.4\,\mu$m range, we repeated the analysis described in Sect.\,\ref{subsec:methods} using data from the GALACTICNUCLEUS survey \citep{NoguerasLara2018,NoguerasLara2019} that spatially overlaps with our JWST observations. The overlapping region is shown in Fig.\,\ref{mosaic_ex}. GALACTICNUCLEUS uses observations taken with the HAWK-I instrument at the VLT \citep{KisslerPatig2008}. We selected these data because they are the most complete ones after JWST, thanks to their angular resolution, and because the GALACTICNUCLEUS JH$\mathrm{K_s}$ filters overlap with our JWST filters—specifically, J with F115W, H with F162M, and $\mathrm{K_s}$ with F212N. Due to the limited number of detections in the GALACTICNUCLEUS J band in this region (only $\sim$1700 stars), we restricted our analysis to the H and $\mathrm{K_s}$ bands.

We retrieved the GALACTICNUCLEUS data from the ESO Archive\footnote{\url{https://archive.eso.org/scienceportal/home}}. Figure\,\ref{app_hawki} shows the RC selection box and binning. We computed the extinction ratios for both RC components, along with their uncertainties, and derived the corresponding extinction indices using Eq.\,\ref{alpha_calc}. By combining the statistical and systematic uncertainties in quadrature, and averaging the results from both RC components, we obtained a final extinction index of $\alpha_\mathrm{HK_s} = 2.13 \pm 0.10$. This result is consistent with the value reported by \citet{NoguerasLara2020}, $\alpha_\mathrm{HK_s} = 2.23 \pm 0.05$, but it differs by about one sigma from the value obtained with our JWST data, $\alpha_\mathrm{F212N-F162M} = 1.92 \pm 0.09$. It is important to note that the region we analysed from GALACTICNUCLEUS contains $\sim 6000$ RC stars, whereas \citet{NoguerasLara2020} used more than 165\,000 RC stars in a much larger field, which may explain the smaller uncertainty of their index.  

\begin{figure}
    \centering
    \includegraphics[width=1\linewidth]{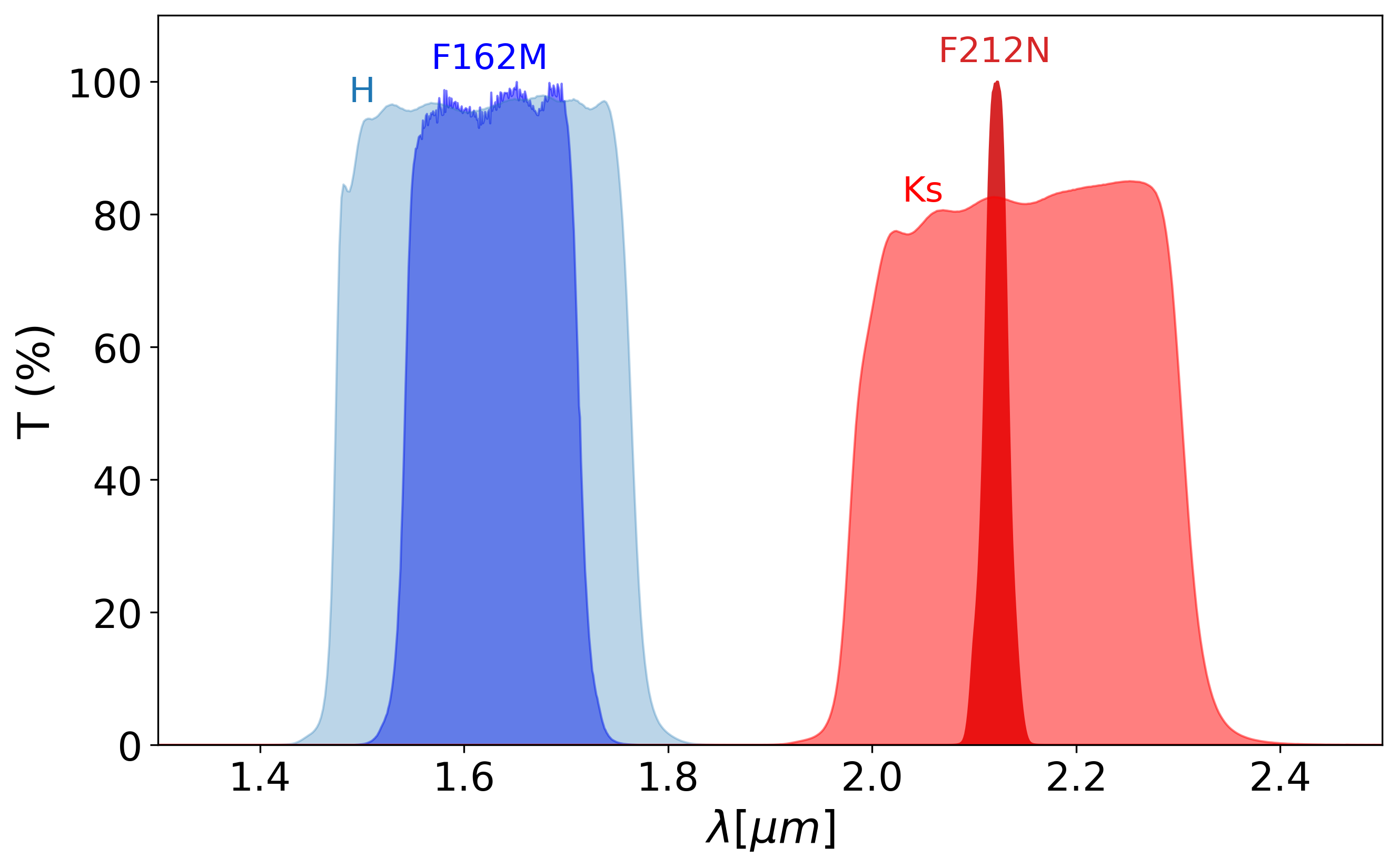}
    \caption{Comparison between the transmission curves of the HAWK-I filters H and $\mathrm{K_s}$ and the JWST filters F162M and F212N.}
    \label{tcs}
\end{figure}

To understand the discrepancy, we compared the transmission curves of the HAWK-I and JWST filters (see Fig.\,\ref{tcs}). The differences are considerable, particularly between the F212N and $\mathrm{K_s}$ filters, with the HAWK-I bands being significantly broader. As a result, the effective wavelengths of the HAWK-I \citep[$\mathrm{H} = 1.6506 \pm 0.0020\,\mu$m, $\mathrm{K_s} = 2.1629\pm 0.0009\,\mu$m; ][]{NoguerasLara2018} and JWST ($\mathrm{F162M} = 1.6342\pm0.0004\,\mu$m, $\mathrm{F212N} = 2.12136\pm 0.00001\,\mu$m) filters are different. If the extinction curve varies significantly within the $\sim$1–2.5\,$\mu$m range \citep[as proposed by][]{Hosek2018, NoguerasLara2019a, NoguerasLara2020}, then even small shifts in effective wavelength can translate into measurable variations in the derived extinction index, which might justify the difference that we found.

This effect is compounded by the fact that broader filters like $\mathrm{K_s}$ will sample a wider portion of the extinction curve, potentially averaging over regions with different slopes. Consequently, the extinction index derived using $\mathrm{K_s}$ may not correspond to a single point on the curve but rather to an average over a broader wavelength range, which can lead to discrepancies when comparing with results from narrower filters like F212N. 

In addition, when using broad-band filters the measured extinction index and effective wavelength will depend on the type of stars observed. As discussed in \citet{NoguerasLara2020}, this can introduce systematic biases. 

Taken together, the combination of differences in filter profiles, the averaging effects of broad-band filters, and the presence of non-linear photometric effects provides a natural explanation for the discrepancy observed between our extinction index and that of the GALACTICNUCLEUS dataset. We cannot confirm the wavelength dependence of the extinction index in the 1–2.5\,$\mu$m range reported by \citet{NoguerasLara2020}. However, considering the differences in the datasets, the targeted regions, and the challenges associated with accurately determining extinction, we cannot exclude this possibility either.

\subsection{Comparison with previous extinction curves} \label{GCcurves}

Numerous infrared extinction curves have been reported in the literature, several of them based on GC data, obtained using different instruments and methodologies—each with its own wavelength coverage. As a result, a range of values for extinction ratios and indices has been published. In this section we compare our derived extinction curve with previous curves obtained both from GC data and from observations targeting other regions within the Galaxy and, in one case, from an extragalactic environment.  

\subsubsection{Extinction curves derived with GC data} \label{GCcurvesGC}

\begin{figure*}
    \centering
    \includegraphics[width=1\linewidth]{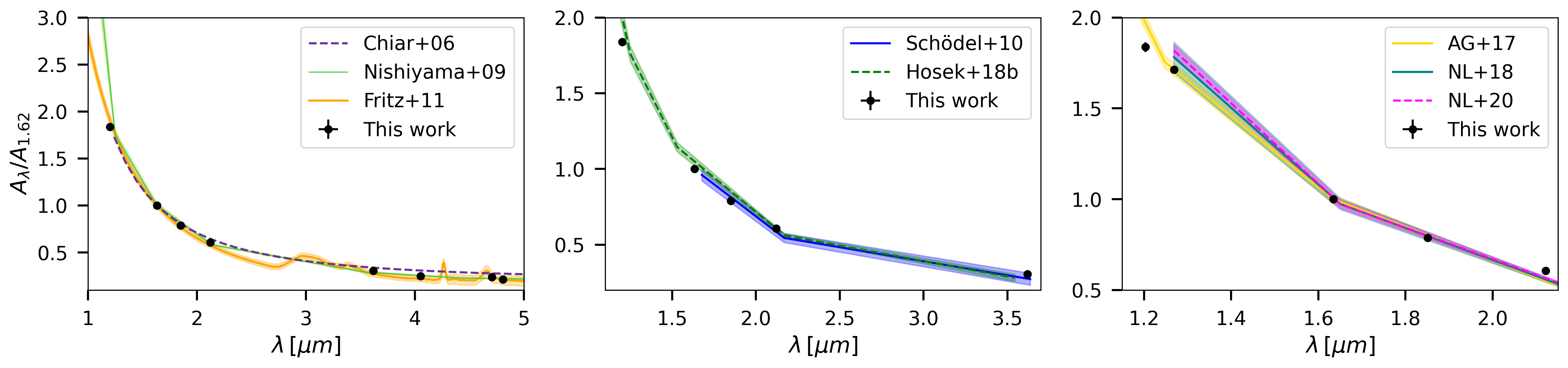}
    \caption{Extinction curve derived in this work compared with previous results from the literature. Left: \citet{Chiar2006}, \citet{Nishiyama2009}, and \citet{Fritz2011}. Centre: \citet{Schoedel2010} and \citet{Hosek2019}. Right: \citet{AlonsoGarcia2017}, \citet{NoguerasLara2018}, and \citet{NoguerasLara2020}. In all cases we translated the extinction curves from the literature to our reference filter system. For the centre and right panels, the J-band extinction value and its uncertainty were obtained by interpolating our extinction curve. Uncertainties are not available for \citet{Chiar2006} and are therefore not shown.}
    \label{comp_lit}
\end{figure*}

\renewcommand{\arraystretch}{1.3}
\setlength{\tabcolsep}{6pt}

\begin{table*}[htbp]
\small
\centering
\caption{Extinction ratios $A_\lambda / A_\mathrm{F162M}$ for the NIRCam/JWST filters, interpolated from literature values.}
\label{ratios_lit}

\begin{tabular}{lccccccc}
\hline \hline
Reference
& F115W
& F182M
& F212N 
& F360M 
& F405N 
& F470N 
& F480M \\
\hline
Chiar+06     & --              & $0.804$ &  $0.646$ & $0.338$ & $0.308$ & $0.278$ & $0.275$           \\
Nishiyama+09 & $1.94 \pm 0.05$ & $0.794 \pm 0.020$ & $0.592 \pm 0.010$ & $0.286 \pm 0.008$ & $0.255 \pm 0.009$ & $0.214 \pm 0.006$ & $0.210 \pm 0.006$ \\
Schödel+10   & --              & $0.80 \pm 0.04$   & $0.58 \pm 0.03$   & $0.27 \pm 0.04$   & --                & --                & --                \\
Fritz+11     & $1.91 \pm 0.06$ & $0.77 \pm 0.03$   & $0.58 \pm 0.03$   & $0.27 \pm 0.03$   & $0.22 \pm 0.03$   & $0.23 \pm 0.05$   & $0.20 \pm 0.05$ \\
AG+17        & $1.91 \pm 0.04$ & $0.823 \pm 0.019$ & $0.563 \pm 0.009$ & --                & --                & --                & --                \\
Hosek+18     & $1.93 \pm 0.06$ & $0.772 \pm 0.018$ & $0.586 \pm 0.011$ & --                & --                & --                & --                \\
NL+18        & --              & $0.72 \pm 0.09$   & $0.545 \pm 0.012$ & --                & --                & --                & --                \\
NL+20        & --              & $0.705 \pm 0.017$ & $0.536 \pm 0.009$ & --                & --                & --                & --                \\
This work & $1.84 \pm 0.03$ & $0.789 \pm 0.005$ & $0.607 \pm 0.014$ & $0.306 \pm 0.011$ & $0.248 \pm 0.017$ & $0.240 \pm 0.019$ & $0.21 \pm 0.03$ \\
\hline
\end{tabular}
\tablefoot{ For comparison, we also included the extinction ratios derived in this work. We added quadratically the statistical and systematic uncertainties.}
\end{table*}

We contrasted our extinction curve (using the reference filter method) with several extinction curves from the literature that are representative of the GC (see Fig.\,\ref{comp_lit}). To express all extinction curves relative to our reference filter, we interpolated them to estimate the extinction at the effective wavelength of our reference filter (F162M). For the interpolation, we used the \texttt{CubicSpline} function from the SciPy library. In addition, Table\,\ref{ratios_lit} shows the extinction ratios interpolated for our filter set based on the literature curves, alongside our results (we added quadratically the statistical and systematic uncertainties). It is important to note that Fig.\,\ref{comp_lit} does not display these interpolated values, but rather the extinction ratios originally reported in the respective works, converted to our reference filter system. The curves appear continuous simply because we connected the reported data points for visualisation purposes.
 
Overall, the trend observed in our extinction curve is consistent with that reported in previous studies: the extinction decreases with increasing wavelength until it reaches small values with local variations due to absorption by different molecules and ices, as clearly seen in Fig.\,\ref{comp_lit}. Based on Table\,\ref{ratios_lit}, we find that the two curves covering the full wavelength range—those of \citet{Nishiyama2009} and \citet{Fritz2011}—are in good agreement with our results, except for $A_\mathrm{F115W}/A_\mathrm{F162M}$ in the case of \citet{Nishiyama2009}.

To reproduce the \citet{Nishiyama2009} extinction curve, we used the data provided in their Table\,1. Their extinction curve was derived photometrically by combining \textit{Spitzer} Infrared Array Camera \citep[IRAC;][]{Fazio2004} observations of bulge fields with Simultaneous-Color InfraRed Imager for Unbiased Survey of the InfraRed Survey Facility \citep[SIRIUS-IRSF;][]{Nagashima1999} data of the GC, incorporating both space- and ground-based measurements. We reproduced the \citet{Fritz2011} curve using their publicly available interpolated infrared extinction curve. This curve was derived from hydrogen emission lines observed with ISO-SWS and SINFONI in the mini-spiral region.

In both cases, the extinction curves correspond to regions that are spatially distinct from our Sgr\,C field: the \citet{Fritz2011} curve was derived in the central mini-spiral, while \citet{Nishiyama2009} analysed a broader area of the central bulge ($|l|\lesssim3^\circ$, $|b|\lesssim1^\circ$). This supports previous evidence that the infrared extinction curve towards the inner Galaxy is not dependent on the line of sight \citep[e.g.][]{NoguerasLara2019}. 

Regarding the extinction index, \citet{Fritz2011} reported a mean value of $\alpha = 2.11 \pm 0.06$ over the wavelength range $1.282$–$2.758\,\mu$m, while \citet{Nishiyama2009} found $\alpha = 2.0$ in the J, H, and $\mathrm{K_s}$ bands. The  $\alpha = 2.0$  is in better agreement with the extinction indices we derived for the SW filters. A plausible explanation for this difference lies in the methodologies used. \citet{Nishiyama2009} derived their extinction curve using a photometric method comparable to ours: they analysed RC stars in near-infrared colour-magnitude and colour-colour diagrams. In contrast, \citet{Fritz2011} used spectroscopic measurements. Thus, both our method and that of \citet{Nishiyama2009} rely on photometry of RC stars, sampling a similar stellar population and being subject to comparable biases—such as filter transmission effects. 

Our results are not consistent with those of \citet{Chiar2006}, as shown in the left panel of Fig.\,\ref{comp_lit} and in Table\,\ref{ratios_lit}. The interpolated extinction ratios are systematically higher than ours. \citet{Chiar2006} combined a photometrically derived continuum extinction curve—based on the works of \citet{Lutz1999} and \citet{Indebetouw2005}—with spectroscopic ISO-SWS data from GCS\,3, an infrared-bright source in the Quintuplet cluster. However, robust conclusions are limited by the lack of reported uncertainties in their extinction values. 

Our extinction curve also shows good agreement with that of \citet{Schoedel2010}, which was derived from photometric observations of the central parsec of the Galaxy. This agreement is evident in both the central panel of Fig.\,\ref{comp_lit} and Table\,\ref{ratios_lit}. Their extinction indices, $\alpha_{\mathrm{H-K_s}} = 2.21 \pm 0.24$ and $\alpha_{\mathrm{K_s-L'}} = 1.34 \pm 0.29$, are consistent—within relatively large uncertainties, which likely stem from significant sources of error such as zero point uncertainties and the limited number of RC stars used (fewer than $\sim7000$)—with our corresponding values: $\alpha_{\mathrm{F162M-F212N}} = 1.92 \pm 0.09$ and $\alpha_{\mathrm{F360M-F212N}} = 1.28 \pm 0.07$. These particular filters were chosen due to their overlap. The first index was computed directly as described in Appendix\,\ref{app1}, while we calculated the second using Eq.\,\ref{alpha_calc}, based on the derived extinction ratios and effective wavelengths. We obtained its associated uncertainty through standard quadratic error propagation.

Regarding the extinction curve of \citet{Hosek2019}, which spans the 0.7–3.54\,$\mu$m range, the authors combined \textit{Hubble} Space Telescope observations of the GC ($l = 0.^{\circ}121$, $b=0.^{\circ}0168$) and Westerlund\,1 ($l = - 20.^{\circ}451$, $b = -0.^{\circ}404$) with VISTA Variables in the Via Lactea \citep[VVV;][]{Minniti2010} $\mathrm{K_s}$-band data for Westerlund\,1, thereby integrating both space- and ground-based datasets and covering different lines of sight. The extinction ratios interpolated from their curve agree with our results within uncertainties, supporting the hypothesis that the infrared extinction curve towards the GC is not dependent on the line of sight.

In contrast, our extinction curve is not compatible with those reported by \citet{AlonsoGarcia2017}, \citet{NoguerasLara2018}, and \citet{NoguerasLara2020}, as shown in the right panel of Fig.\,\ref{comp_lit}. \citet{AlonsoGarcia2017} analysed several regions across the Galactic bulge using broad-band filters (Z, Y, J, H, and $\mathrm{K_s}$) from the VVV survey. \citet{NoguerasLara2018} and \citet{NoguerasLara2020} studied different areas within the GC using GALACTICNUCLEUS data. These studies employed the same methodology as ours to derive the extinction curve.

To enable a more direct comparison, we interpolated our extinction curve to estimate the extinction at the J band using the effective wavelength computed in \citet{NoguerasLara2018}. For the interpolation, we used the \texttt{CubicSpline} function. Visually, the value of $A_\mathrm{J}/A_\mathrm{F162M}$ aligns only with the result from \citet{AlonsoGarcia2017}. To further examine this, we computed $A_\mathrm{J}/A_\mathrm{K_s}$ for comparison with values reported in the literature, interpolating our curve to obtain $A_\mathrm{K_s}/A_\mathrm{F162M}$. We derived $A_\mathrm{J}/A_\mathrm{K_s} = 2.84 \pm 0.08$. This is lower than the values reported by \citet[$3.30 \pm 0.04$]{AlonsoGarcia2017}, \citet[$3.41 \pm 0.14$]{NoguerasLara2018}, and \citet[$3.441 \pm 0.023$]{NoguerasLara2020}. Although the value from \citet{AlonsoGarcia2017} appears to be the closest, it still does not agree with our result within the uncertainties. The apparent agreement in Fig.\,\ref{comp_lit} is misleading. Since the authors do not report effective wavelengths, we had to adopt them from the Spanish Virtual Observatory Filter Profile Service\footnote{\url{http://svo2.cab.inta-csic.es/svo/theory/fps/}} \citep{Rodrigo2012, Rodrigo2020}, introducing uncertainty in the x-axis placement of $A_\mathrm{J}/A_\mathrm{F162M}$. Consequently, its true position could differ significantly from what is shown in the plot. The extinction ratio showing the largest disagreement is $A_\mathrm{F212N}/A_\mathrm{F162M}$, which supports the idea of a wavelength-dependent extinction curve. The broad-band filters used in those studies are significantly wider than F212N, which is a narrow-band filter (see Fig.\,\ref{tcs}).

Regarding the extinction indices, \citet{AlonsoGarcia2017} found $\alpha=2.47 \pm 0.11$ over their full wavelength range. \citet{NoguerasLara2018} reported $\alpha = 2.30 \pm 0.08$ for the $\mathrm{JHK_s}$ range, and \citet{NoguerasLara2020} found two values: $\alpha_\mathrm{JH} = 2.44 \pm 0.05$ and $\alpha_\mathrm{HK_s} = 2.23 \pm 0.05$. All of these are larger than the extinction indices we derived from the SW filters. To make decisive progress, we propose  mapping the extinction curve in the GC systematically with narrow-band filters, from the space and from the ground and with space-based spectroscopy (NIRSpec and MIRI). 

\subsubsection{Other extinction curves}

\begin{figure*}
    \centering
    \includegraphics[width=1\linewidth]{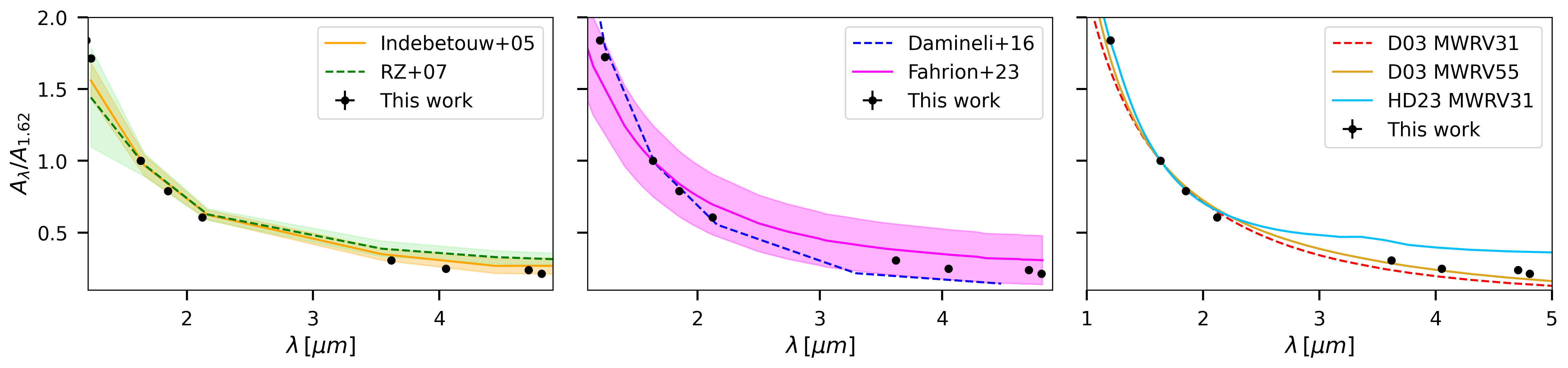}
    \caption{Extinction curve derived in this work compared with extinction curves from the literature derived outside the GC. Left: \citet{Indebetouw2005} and \citet{RomanZuniga2007}. Centre: \citet{Damineli2016} and \citet{Fahrion2023}. Right: Dust grain models from \citet{Draine2003} for $R_V = 3.1$ and $R_V = 5.5$, and from \citet{Hensley2023}, generated using the \texttt{dust\_extinction} package \citep{Gordon2024b}. We translated the extinction curves to our reference filter system. In the left and centre panels, the J-band extinction value and its uncertainty were obtained by interpolating our extinction curve. Uncertainties are not available for \citet{Damineli2016} and are therefore not shown.}
    \label{comp_lit_out}
\end{figure*}

\renewcommand{\arraystretch}{1.3}
\setlength{\tabcolsep}{6pt}

\begin{table*}[htbp]
\small
\centering
\caption{Interpolated extinction ratios $A_\lambda / A_\mathrm{F162M}$ for different extinction curves from the literature derived in regions outside the GC.}
\label{ratios_out}

\begin{tabular}{lccccccc}
\hline \hline
Reference
& F115W
& F182M 
& F212N
& F360M 
& F405N 
& F470N
& F480M \\
\hline
Indebetouw+05 & --              & $0.80 \pm 0.05$ & $0.64 \pm 0.03$ & $0.34 \pm 0.04$ & $0.30 \pm 0.05$ & $0.26 \pm 0.05$              & $0.26\pm0.05$             \\
RZ+07         & --              & $0.82 \pm 0.05$ & $0.65 \pm 0.04$ & $0.38 \pm 0.05$ & $0.36 \pm 0.05$ & $0.32 \pm 0.04$ & $0.31 \pm 0.04$ \\
Damineli+16   & $1.955$         & $0.790$           & $0.579$           & $0.189$           & $0.163$           & --              & --              \\
Fahrion+23    & $1.6 \pm 0.3$ & $0.84 \pm 0.23$ & $0.70 \pm 0.21$ & $0.38 \pm 0.18$ & $0.35 \pm 0.17$ & $0.31 \pm 0.17$ & $0.31 \pm 0.17$ \\
This work
& $1.84 \pm 0.03$ 
& $0.789 \pm 0.005$ 
& $0.607 \pm 0.014$ 
& $0.306 \pm 0.011$ 
& $0.248 \pm 0.017$ 
& $0.240 \pm 0.019$ 
& $0.21 \pm 0.03$ \\
\hline
\end{tabular}
\tablefoot{ For comparison, we also included the extinction ratios derived in this work. We added quadratically the statistical and systematic uncertainties.}
\end{table*}

To compare our extinction curve with others derived from regions outside the GC, we selected the extinction curves reported by \citet{Indebetouw2005}, \citet{RomanZuniga2007}, \citet{Damineli2016}, and \citet{Fahrion2023}, as they represent a variety of environments. \citet{Indebetouw2005} derived their extinction curve using the Two Micron All Sky Survey \citep[2MASS;][]{Skrutskie2006} $\mathrm{JHK_s}$ photometry and \textit{Spitzer} data in the 3–8\,$\mu$m range, targeting two lines of sight: $l = 42^\circ$ (a quiescent region) and $l = 284^\circ$ (the giant H\,II region RCW 49). \citet{RomanZuniga2007} obtained their extinction curve for the 1.25–7.76\,$\mu$m range using ground-based observations with ESO New Technology Telescope (NTT) and VLT, as well as \textit{Spitzer}-IRAC data, focusing on Barnard 59, a dense, star-forming core in the Pipe Nebula. \citet{Damineli2016} combined photometric and spectroscopic data from multiple instruments to derive the extinction curve for Westerlund 1 over the 0.4–4.8\,$\mu$m range. Finally, \citet{Fahrion2023} derived the extinction curve for 30 Doradus in the Large Magellanic Cloud using JWST-NIRCam observations in the 0.3–4.8\,$\mu$m range.

The resulting comparison is shown in Fig.\,\ref{comp_lit_out}, where our extinction curve is plotted alongside those from the literature and several theoretical dust models. For the left and centre panels, we included the J-band extinction ratio by interpolating our extinction curve. For the left panel, we adopted the isophotal wavelengths of 2MASS and IRAC filters convolved with a K2 III stellar spectrum, following the approach in \citet{Indebetouw2005} and \citet{RomanZuniga2007}. For the right panel, we used the J-band effective wavelength provided by \citet{Damineli2016}. The dust grain models shown in the figure are included for reference only, and a detailed comparison is beyond the scope of this work.

Table\,\ref{ratios_out} presents the interpolated extinction ratios for the literature extinction curves, alongside those derived in this work, allowing for a more direct comparison. As seen in both the left panel of Fig.\,\ref{ratios_out} and the table, our results are consistent with those reported by \citet{Indebetouw2005}. This supports the hypothesis that the infrared extinction curve is independent of the line of sight, as their curve was derived from two distinct Galactic directions. Our extinction ratios also match those reported by \citet{RomanZuniga2007}, although only in the SW regime. Both studies used \textit{Spitzer}-IRAC data in the LW range; the discrepancies with our results may be due to the broad \textit{Spitzer} filter profiles and the fact that Barnard 59 is a cold cloud, where the presence of ices can significantly affect extinction in this regime \citep{Riaz2009, Chu2020, Chu2021}. Additionally, we computed the $A_\mathrm{J}/A_\mathrm{K_s}$ ratio using our extinction curve, as described in Sect.\,\ref{GCcurvesGC}, obtaining a value of $A_\mathrm{J}/A_\mathrm{K_s} = 2.97 \pm 0.09$. This is higher than the values reported by \citet[$2.5 \pm 0.2$]{Indebetouw2005} and \citet[$2.3 \pm 0.5$]{RomanZuniga2007}.

Our extinction curve also shows agreement with that reported by \citet{Damineli2016} in the SW regime—except for the $A_\mathrm{F115W}/A_\mathrm{F162M}$ ratio—but deviates in the LW regime, as illustrated in the right panel of Fig.\,\ref{comp_lit_out} and in Table\,\ref{ratios_out}. However, a robust comparison is limited by the fact that \citet{Damineli2016} do not provide uncertainties for their extinction ratios. Unlike the B59 case, this discrepancy is unlikely to be driven by ice absorption, as the line of sight towards Westerlund\,1 shows little evidence of ice \citep{Guarcello2025}. Similarly, our extinction curve agrees with that of \citet{Fahrion2023}, but the associated uncertainties are too large for firm conclusions to be drawn. 

While detailed modelling of the dust properties in Sgr\,C is beyond the scope of this work, we note that the theoretical curve corresponding to $R_V = 5.5$ from \citet{Draine2003} appears to be a closer match to our data than the $R_V = 3.1$ model. This is consistent with expectations for denser environments with larger grains. Our results differ from the \citet{Hensley2023} model in the LW regime. However, we caution that our extinction curve is derived from a localised region, and no strong conclusions about dust grain properties or $R_V$ can be drawn.

In summary, our extinction curve follows the same general trend as the selected literature curves derived outside the GC. It shows excellent agreement with the curve by \citet{Indebetouw2005}, reinforcing the idea of a broadly line-of-sight--independent infrared extinction curve. It is also consistent with the curve by \citet{Fahrion2023}, although due to the large uncertainties we cannot perform a definitive assessment.

\section{Conclusions}

In this work we aimed to determine the extinction curve towards the Sgr\,C region using JWST-NIRCam data, and to evaluate its consistency with previous results. We applied two different methods based on the slopes of the RC features to derive the extinction ratios. The results are shown in Tables \ref{results_short} and \ref{results_long}. 

We find that the extinction curve flattens in the LW regime (3.6–4.8\,$\mu$m), in agreement with previous literature. Although we obtained different extinction indices for the SW and LW regimes, we found no significant wavelength dependence within either regime. However, using GALACTICNUCLEUS data overlapping with our JWST field, we derived an H–$\mathrm{K_s}$ extinction index consistent with the results of \citet{NoguerasLara2020} but not with our own results. We attribute this discrepancy mainly to the difference in filter widths: the H and $\mathrm{K_s}$ filters are much broader than our F162M and F212N filters, leading to significantly different transmission curves. Additionally, these wide-band filters are affected by non-linear effects. 

Furthermore, our extinction curve is in good agreement with those of \citet{Fritz2011}, \citet{Nishiyama2009}, \citet{Schoedel2010}, and \citet{Hosek2019}, which were derived for different lines of sight in the GC. This suggests that the extinction index in the inner Galaxy does not vary significantly between different lines of sight within the uncertainties. In contrast, our extinction ratios and indices are not consistent with the steeper curves reported by \citet{AlonsoGarcia2017}, \citet{NoguerasLara2018}, and \citet{NoguerasLara2020}. We conclude that this discrepancy is most likely due to differences in the filter sets used: in those three studies, the filters are considerably broader than those used here.

We also compared our extinction curve with curves derived from regions outside the GC. We find good agreement with the curve of \citet{Indebetouw2005}, supporting the idea of a largely spatially invariant infrared extinction curve, even beyond the GC. Additionally, our extinction curve agrees with those of \citet{RomanZuniga2007} and \citet{Damineli2016}, but only in the SW regime. Moreover, it is consistent with the curve reported by \citet{Fahrion2023} for 30 Doradus, although the uncertainties in that study are too large for definitive conclusions to be drawn.

\begin{acknowledgements}
This work is based on observations made with NASA/ESA/CSA James Webb Space Telescope. The data were obtained from the Mikulski Archive for Space Telescopes at the Space Telescope Science Institute, which is operated by the Association of Universities for Research in Astronomy, Inc., under NASA contract NAS 5-03127 for JWST. These observations are associated with program 4147. The specific observations analysed can be accessed via \url{https://doi.org/10.17909/pwkh-ec13}.

LBF and RS acknowledge support from grant PID2022-661136640NB-C21, funded by MCIN/AEI/10.13039/501100011033 and by the European Union. FNL acknowledges support from grant PID2024-162148NA-I00, funded by MCIN/AEI/10.13039/501100011033 and the European Regional Development Fund (ERDF) “A way of making Europe”. LBF, FNL, and RS acknowledge support from the Severo Ochoa grant CEX2021-001131-S, funded by MCIN/AEI/10.13039/501100011033. R.F. acknowledges support from the grants PID2023-146295NB-I00, and from the Severo Ochoa grant CEX2021-001131-S funded by MCIN/AEI/ 10.13039/501100011033 and by ``European Union NextGenerationEU/PRTR''. Y.C. was partially supported by a Grant-in-Aid for Scientific Research (KAKENHI  number JP24K17103) of the JSPS. ZYL is supported in part by NASA 80NSSC20K0533 and NSF AST-2307199.
\end{acknowledgements}

  \bibliographystyle{aa} 
  \bibliography{paperI}

\onecolumn
\begin{appendix} 

\section{Effective wavelength} \label{eff_wls_sect}

We computed the effective wavelengths following the approach described in Appendix\,B of \citet{NoguerasLara2018}. That is, we applied Eq.\,A3 from \citet{Tokunaga2005}. We used the transmission curves for the JWST-NIRCam filters (see Sect.\,\ref{data}), available on the instrument website\footnote{\url{https://jwst-docs.stsci.edu/jwst-near-infrared-camera/nircam-instrumentation/nircam-filters}}. We computed $\lambda_{\mathrm{eff}}$ for RC stars using the most adequate Kurucz model. We adopted the nuclear stellar disc west value of $A_\mathrm{1.61}$ from \cite{NoguerasLara2020}, as Sgr\,C is located at the western edge of the GC. To calculate $\alpha$ we used an iterative approach. Namely, we computed a first value for $\alpha$ using the relation

    \begin{equation}
        \frac{A_\mathrm{\lambda_1}}{A_\mathrm{\lambda_2}} = \left(\frac{\lambda_1}{\lambda_2}\right)^{\alpha}
        \label{alpha_calc}
    \end{equation}

    \noindent from the extinction ratios and a first estimate of $\lambda$ for each filter. We then computed $\lambda_{\mathrm{eff}}$ and recalculated $\alpha$, iterating this process updating until we reached convergence. 
    
Uncertainties in $\lambda_{\mathrm{eff}}$ were estimated by varying the relevant parameters within their uncertainty ranges, changing one at a time while keeping the others fixed. We considered the following parameters:

    \begin{itemize}
        \item The effective temperature of the RC star model: 4500, 4750, and 5000 K.
        \item The metallicity of the model: 0.0, 0.3, and 0.5 dex.
        \item The extinction index, according to its uncertainty, derived from the uncertainty of the extinction ratio. 
        \item The value of $A_{1.61}$, using the uncertainty given by \citet[$2.96 \pm 0.12$]{NoguerasLara2020}.
    \end{itemize}

The final uncertainty in $\lambda_{\mathrm{eff}}$ was computed by adding quadratically all individual contributions. The final values of $\lambda_{\mathrm{eff}}$ were calculated using the central values of temperature and metallicity, and are presented in Table\,\ref{eff_wls}. As expected, the uncertainties for the narrow-band filters are significantly smaller—one to two orders of magnitude lower—since these filters cover a narrower wavelength range.

\setlength{\tabcolsep}{10pt}
\begin{table}[h]
    \centering
    \caption{Obtained values for the extinction index and the effective wavelengths using extinction ratios with the reference filter method.}
    \label{eff_wls}
    \vspace{0.5em}
    
    \begin{tabular}{l l}
    \hline \hline
        
        \multicolumn{1}{l}{ Filter}    & \multicolumn{1}{l}{$\lambda_{\mathrm{eff}}$ ($\mu$m)} \\ \hline
        \multicolumn{1}{l}{F115W}  & $1.2030 \pm 0.0021$              \\ 
        \multicolumn{1}{l}{F162M}  & $1.6342 \pm 0.0004$              \\ 
        \multicolumn{1}{l}{F182M}  & $1.8509 \pm 0.0005$              \\ 
        \multicolumn{1}{l}{F212N}  & $2.12136 \pm 0.00001$              \\ 
        \multicolumn{1}{l}{F360M}  & $3.61967 \pm 0.00022$              \\ 
        \multicolumn{1}{l}{F405N}  & $4.051621 \pm 0.000004$              \\ 
        \multicolumn{1}{l}{F470N}  & $4.70774 \pm 0.00003$              \\ 
        \multicolumn{1}{l}{F480M}  & $4.8095 \pm 0.0003$              \\ \hline
    \end{tabular}

\end{table}

\FloatBarrier
\section{RC selection boxes} \label{app1}

\begin{figure*}[t]
        \centering
        \includegraphics[width=0.95\linewidth]{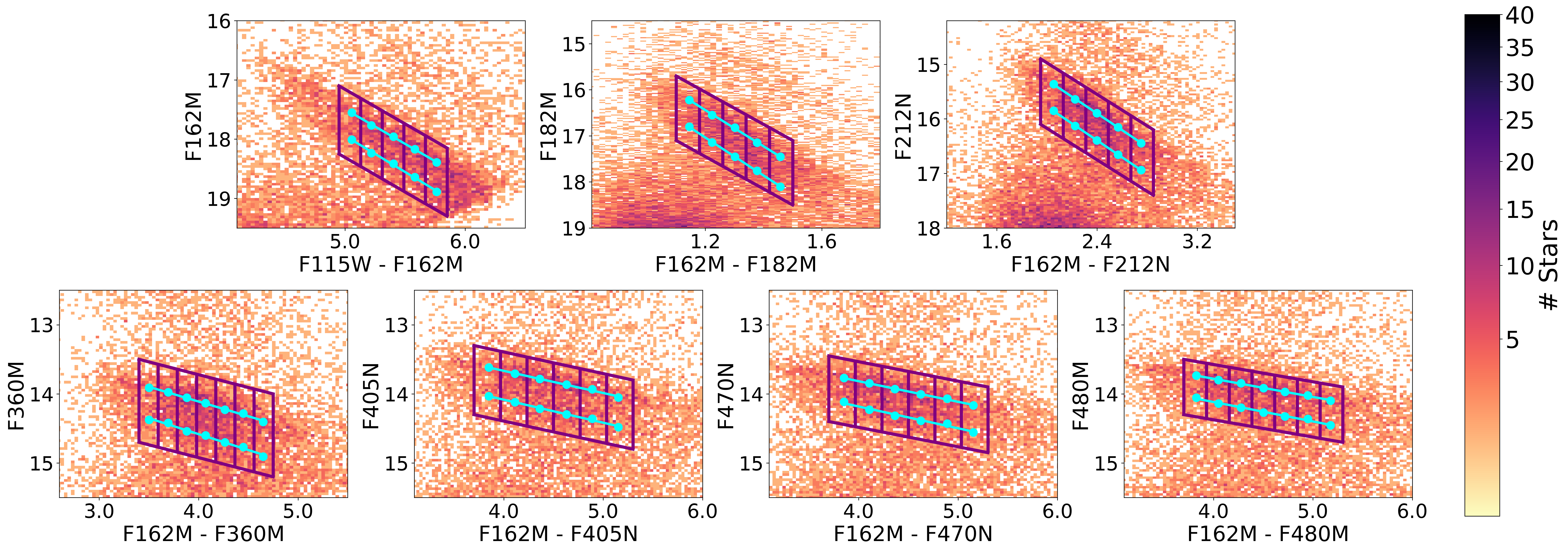}
        \caption{Selection boxes and bins for each CMD for the reference filter method, shown in purple. Cyan points indicate the results obtained from the GMM with their associated uncertainties. Cyan lines represent the linear fits to these points.}
        \label{app_method1}
    \end{figure*}

\begin{figure*}[t]
        \centering
        \includegraphics[width=0.8\linewidth]{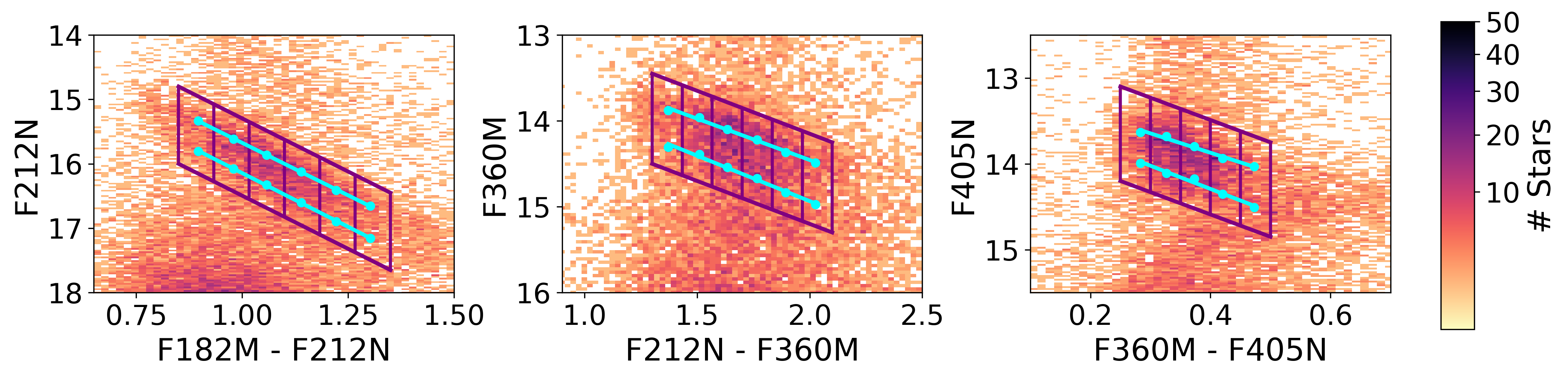}
        \caption{Same as Fig. \ref{app_method1} but for the consecutive filter method.}
        \label{app_method2}
    \end{figure*}

Figures\,\ref{app_method1} and \ref{app_method2} show the RC selection boxes for the reference and consecutive filter methods, respectively, along with the bins defined for each CMD and the two RC features identified with the GMM. Note that the F162M versus F115W-F162M and F182M versus F162M-F182M CMDs are identical in the two methods. 
    
    \begin{figure*}[t]
        \centering
        \includegraphics[width=0.3\linewidth]{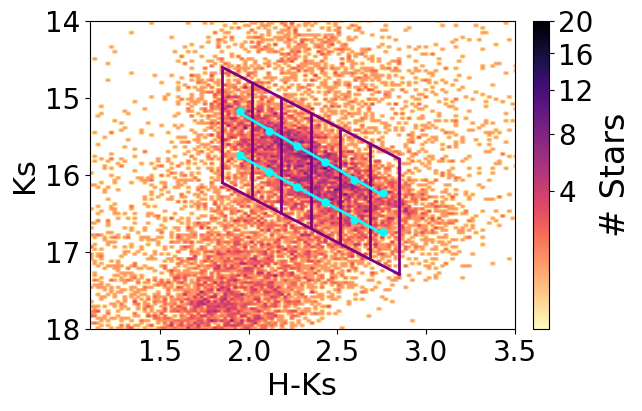}
        \caption{Selection box and bins for the $\mathrm{K_s}$ versus H-$\mathrm{K_s}$ CMD from GALACTICNUCLEUS, shown in purple. Cyan points indicate the results obtained from the GMM with their associated uncertainties. Cyan lines represent the linear fits to these points}
        \label{app_hawki}
    \end{figure*}

\section{Foreground population in LW filter CMDs} \label{app2}

\begin{figure}
    \centering
    \includegraphics[width=0.6\linewidth]{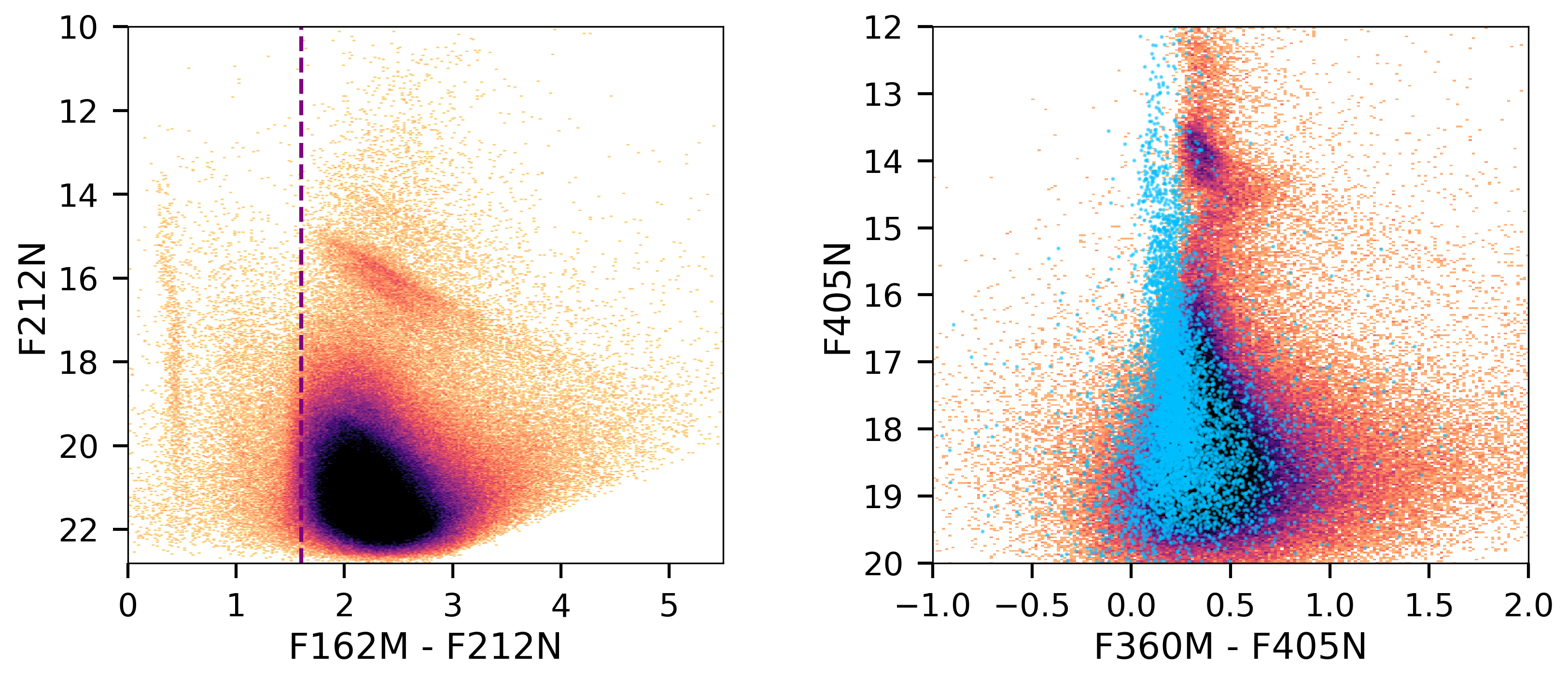}
    \caption{Top: F212N versus F162M-F212N CMD. The dashed purple line indicates that stars with F162M-F212N $< 1.6$ belong to the foreground population. Bottom: F405N versus F360M-F405N CMD. We cut the CMD at F405N $= 12$\,mag because that is the saturation limit. The blue points are the foreground stars identified in the F212N versus F162M-F212N CMD.}
    \label{fore_id}
\end{figure}

Figure\,\ref{fore_id} shows the foreground population selection in the F212N versus F162M-F212N CMD (top panel) along with the F405N versus F360M-F405N CMD with the foreground stars in blue (bottom panel). We cut the latter CMD at F405N = 12 mag because that is the saturation limit.

\end{appendix}

\end{document}